\documentclass[%
 aip,
rsi,%
 amsmath,amssymb,
 reprint,%
floatfix 
]{revtex4-1}

\usepackage{graphicx}
\usepackage{dcolumn}

\usepackage{bm} 
\usepackage{amsmath}  
\usepackage{amsfonts} 
\usepackage{graphicx} 
\usepackage[titletoc,title]{appendix}

\newcommand{\be}{\begin{equation}}
\newcommand{\bec}{\begin{center}}
\newcommand{\eec}{\end{center}}
\newcommand{\ee}{\end{equation}}
\usepackage{multirow}

\begin{document}

\title{Particle Plasmons: Why Shape Matters}

\author{William L. Barnes}
\email{w.l.barnes@exeter.ac.uk}
\affiliation{Department of Physics and Astronomy,  University of Exeter, Stocker Road, 
 Exeter, EX4 4QL, United Kingdom}

\begin{abstract}
Simple analytic expressions for the polarizability of metallic nanoparticles are in wide use in the field of plasmonics, but the physical origins of the different terms that make up the polarizability are not obvious.  In this article expressions for the polarizability of a particle are derived in the quasistatic limit in a way that allows the physical origin of the terms to be clearly seen.  The discussion is tutorial in nature, with particular attention given to the role of particle shape since this is a controlling factor in particle plasmon resonances.
\end{abstract}

\maketitle

\section{Introduction} 

\begin{figure}[tbp] 
\begin{center}
\mbox{\scalebox{0.5}{%
\includegraphics{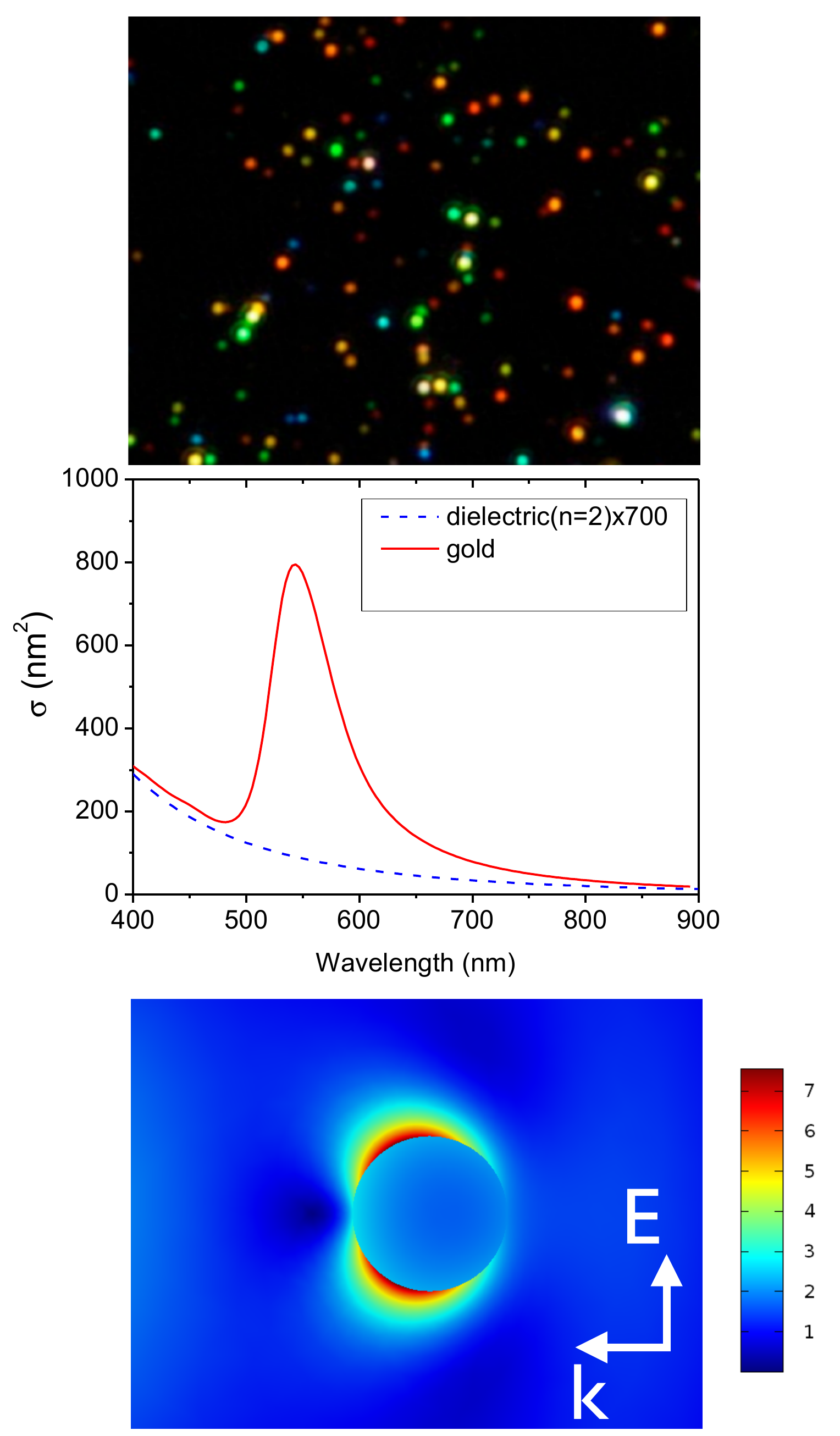}}} 
\caption{Upper. A dark-field microscope image of a suspension of gold and silver nanoparticles, courtesy of the Mulvaney research group, Melbourne. The colour of the particles depends on their size, shape, environment and composition. Middle. Calculated scattering cross-sections for $\mathrm{TiO_2}$ (refractive index assumed to be 2.0) and Au particles, both 100 nm in diameter, in water. Note that the cross-section for the $\mathrm{TiO_2}$ particle has been multiplied by $700$. Lower. Time-averaged electric field strength around a 100 nm gold nanosphere in water, the particle is illuminated on resonance (wavelength = 550 nm) from the right by an electromagnetic plane wave. The colour bar shows the strength of the electric field relative to the incident field, in the plane that contains the particle centre and the direction of the incident field. Note how the field is tightly confined to the vicinity of the particle (central circular region). The asymmetry is a result of the size of the sphere (100 nm) being too big for the quasistatic approximation to hold. The field distribution was calculated numerically using COMSOL\texttrademark. The minimum relative field strength is 0.0, the maximum is 7.55. The relative permittivity for these calculations was based on. \cite{LandH}}
\label{fig:Figure_1}
\end{center}
\end{figure}

A flu virus cannot be seen by eye, even with the best optical microscopes but, amazingly, a metal particle of the same size ($\sim$100 nm) can be seen with ease, how so?  Light impinging on such a particle sets the electrons within it into a ÔringingÕ motion: this ringing mode, known as a plasmon mode, is at the heart of a field of study called plasmonics.  Just as a ringing bell has a particular note, the ringing electrons scatter light of a particular colour, the specific colour depending on the size and shape of the particle, and the particle's immediate environment.  Classical and medieval craftsmen exploited this effect unwittingly; the colour in some medieval stained glass arises from light scattered by metallic nanoparticles, particles that formed from impurities when the glass cooled; a beautiful example is the Lycurgus cup, a legacy of the late Roman period.\cite{Barber_Archaeometry_1990_32_33} In more recent times, the different colours that arise when light is scattered by gold colloids suspended in water was keenly studied by Faraday,\cite{Faraday_PhilTrans_1857_147_145} and understood through the application of electrodynamics by Mie, \cite{Mie_AnnPhys_1908_25_377} see Fig. \ref{fig:Figure_1}.

In addition to scattering, the interaction between light and metallic nanostructures can lead to the concentration of light into sub-wavelength volumes. \cite{Gramotnev_NatPhot_2014_8_13,Murray_NL_2006_6_1772}  It is this control over light deep into the sub-wavelength (nanometre) regime that is behind much of the recent excitement in the field of plasmonics.  In addition to light being confined, the strength of the light is also enhanced, by up to several orders of magnitude.   This enhancement of the (electric) field associated with the light is at the heart of phenomena such as: surface enhanced Raman scattering (SERS), \cite{LRandE} plasmon-mediated absorption  \cite{Andrew_JMO_1997_44_395} and emission \cite{Gruhlke_PRL_1986_56_2838, Kitson_PRB_1995_52_11441} of light; plasmon-mediated strong coupling; \cite{Torma_RepProgPhys_2015_78_013901} THz generation \cite{Polyushkin_NL_2011_11_4718} and some forms of scanning probe microscopy. \cite{Fischer_PRL_1989_62_458} Scattering is being pursued as a means to enhance the efficiency of some types of solar cells, \cite{Atwater_NatMat_2010_9_205} whilst absorption is being exploited for its potential as a photo-thermal treatment for cancer. \cite{Loo_NL_2005_5_709} Particle plasmon dominated extinction, the combination of absorption and scattering, is a powerful and maturing tool for bio-sensing. \cite{Willets_AnnuRevPhysChem_2007_58_267} Silver nanoparticles are even being considered as a means to simultaneously imbue fabrics with antibacterial properties and color via plasmon modes. \cite{Wu_AdvFuncMat_2015}

Although it has a long history, plasmonics is still rapidly expanding, for example the use of plasmonic \textquoteleft atoms\textquoteright\ in metamaterials. \cite{Meinzer_NatPhot_2014_8_889}  Those new to the field may find it difficult to assimilate the latest results whilst at the same time trying to develop a deep understanding of the fundamentals.  This tutorial-style article provides a starting point by looking at the simplest description of the plasmonic response of metallic nanoparticles - one based upon the quasistatic polarizability.

The polarizability determines how strongly a particle scatters and absorbs light, and the degree to which the incident field is enhanced in the vicinity of the particle.  Scattering (see Fig. \ref {fig:Figure_1} (middle)) and absorption spectra are calculated from the polarizability, $\alpha$, using,

\be
\label{eq:num_sig_scatt}
\sigma_{sca} = \frac{k_0^4|\alpha|^2}{6\pi\varepsilon_0^2}, 
\ee

\be
\label{eq:num_abs_scatt}
\sigma_{ext}=\frac{k_0}{\varepsilon_0\sqrt{\varepsilon_1}} \Im(\alpha).
\ee

\noindent
where $\sigma_{sca}$ and $\sigma_{ext}$ are the scattering and extinction cross-sections, $k_0$ is the free-space wavevector, $\varepsilon_0$ is the permittivity of free space and $\varepsilon_1$ is the relative permittivity of the medium in which the particle is embedded, and $\Im(\alpha)$ is the imaginary part of the polarizability. The formula for the scattering cross-section will be derived below.  From the extinction cross section of Ref~\citenum{LRandE} (Eq.~5.35), the absorption cross-section is given by $\sigma_{abs}=\sigma_{ext}-\sigma_{sca}$.

In the quasistatic approach the analysis is carried out in the static (DC) regime but the material parameters are taken to be frequency dependent. Perhaps surprisingly, this approach is found to be applicable in many experimental situations, even at optical frequencies (see ~\cite{LRandE}, section 5.1.4).  However, care needs to be taken over the derivation of the polarizability if we are to appreciate the physical origin of the different terms involved; an example may illustrate the point.  The polarizability, $\alpha$, of a small spherical object of relative permittivity $\varepsilon$, surrounded by free space, is given by Le Ru and Etchegoin (see ~\cite{LRandE} eqn 6.17) as,
\be
\label{eq:num_1}
\alpha = 4\varepsilon_0\pi R^3\left(\frac {\varepsilon-1}{\varepsilon+2}\right),
\ee
\noindent where $R$ is the radius of the sphere. One can see from equation \eqref{eq:num_1} that a resonance will occur if the denominator goes to zero, i.e. if $\varepsilon = -2$ (often known as the Fr\"olich mode, see ~\cite{BandH} p 327).  When the object is metallic such a resonance is known as a particle plasmon resonance (often referred to as a localized surface plasmon resonance). Why does resonance occur at $\varepsilon = -2$?  One of the main purposes of this article is to answer this question. The approach adopted here is somewhat less than conventional, though not entirely new. \cite{Jones_PR_1945_68_93}

The polarizability is one measure of how easily the charge within an object may be displaced by an applied electric field.  The low mass of conduction electrons (compared to the more massive ion cores) means that in metals we usually consider the conduction electrons to be free to move against a fixed background of net positive charge. The conduction electrons move in response to an external field, producing a net positive charge on one side of the particle and a net negative charge on the other. This displacement of negative and positive charge means that a dipole moment is created, the strength of which depends on the polarizability.

Resonance occurs due to the restoring force (Coulomb attraction) between the displaced positive and negative charges, the strength of the force determines the frequency of the resonance whilst the damping determines the width of the resonance.  Here the restoring force is related to the electric field \textit{inside} the particle that arises due to the displaced charges. The field \textit{outside} the particle does not act on charges \textit{within} the particle, it is the motion of charges within the particle that is of interest here.

For an infinite slab the accumulation of electrons on one surface produces a surface charge density $-S$, given by $-S=-nex$, $n$ being the number density of (displaced) electrons, $e$ the magnitude of their charge and $x$ the displacement. The deficit of electrons on the other surface produces a surface charge density $+S$. If we consider the motion of one electron under the influence of an external applied electric field $E$ then, using Newton's second law,
\be
\label{eq:num_a}
m_e\frac {d^2x}{dt^2}=-eE,
\ee
\noindent where $m_e$ is the mass of the electron. The resonance condition is given when the external applied field is set to zero, in which case the field is solely that due to the polarization charge, i.e. $E_{pol}$. The strength of the field $E_{pol}$ depends on (i) the charge density on the surface and (ii) the shape of the surface. For a planar surface $E_{pol}=S/{\varepsilon_0}$ (see  ~\cite{GriffithsEM} sec 4.2.2). For a non-planar surface the field will be different by some factor, let us call it $L$. For a sphere $L=1/3$ (for a slab $L=1$). We can now write equation \eqref{eq:num_a} in terms of the surface charge density $S$, and hence in terms of the displacement $x$,
\be
\label{eq:num_b}
m_e\frac {d^2x}{dt^2}=-\frac{eLS}{\varepsilon_0}=-\frac{eLnex}{\varepsilon_0}.
\ee
\noindent If we now assume a time dependence of the form $e^{-i\omega t}$, equation \eqref{eq:num_b} leads to a  resonance frequency, $\omega_{res}$,
\be
\label{eq:num_c}
\omega_{res}=\sqrt{\frac{e^2nL}{m_e\varepsilon_0}},
\ee
\noindent for a slab, $L=1$ and we recover the standard result for the plasma frequency, $\omega_{P}$,
\be
\label{eq:num_d}
\omega_{res}=\omega_{P}=\sqrt{\frac{e^2n}{m_e\varepsilon_0}}.
\ee
\noindent For a sphere $L=1/3$ and we find the  resonance frequency to be,
\be
\label{eq:num_e}
\omega_{res}=\frac{\omega_P}{\sqrt{3}}.
\ee

\begin{figure}[htbp] 
\begin{center}
\mbox{\scalebox{0.45}{%
\includegraphics{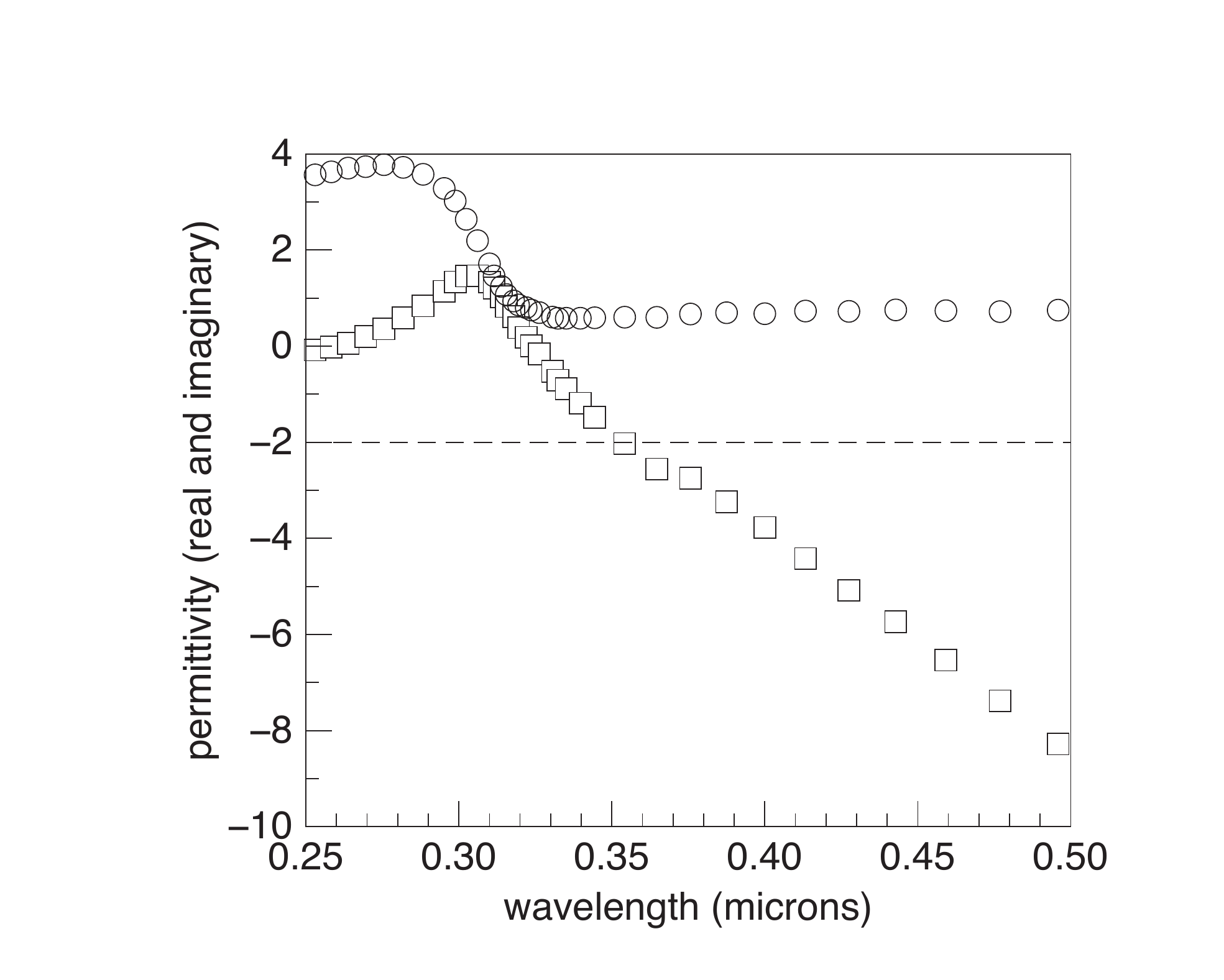}}} 
\caption{The complex permittivity of silver as a function of wavelength. The real and imaginary parts, $\Re({\varepsilon})$ (open squares) and $\Im({\varepsilon})$ (open circles) are shown. For a sphere in free space resonance occurs at $\Re({\varepsilon})=-2$, indicated by a dashed line. Data are taken from. \cite{LandH} It is also possible for the response of the material to be dominated by holes rather than electrons, particle plasmon resonances still occur. \cite{Luther_NatMat_2011_10_361}} 
\label{fig:Figure_2}
\end{center}
\end{figure}

We can estimate the  resonance frequency for a metal such as gold; the free-electron density of gold is $\sim 10^{28}$ $m^{-3}$ so that, using equation \eqref{eq:num_d} the  resonance frequency is $\sim 10^{16} \text{Hz}$, i.e UV/visible. The resonance condition given by equation \eqref{eq:num_1}, can not be met in the static limit, i.e. at zero frequency.  The DC response of dielectrics yields a positive value for the relative permittivity (dielectric constant) i.e. $\varepsilon>1$ whilst that of metals gives $\varepsilon\sim-i\infty$. Meeting the resonance condition is a matter of looking for a dynamic resonant response, i.e. finding a frequency for which the denominator of equation \eqref{eq:num_1} is minimised, see Fig.  \ref {fig:Figure_2}. 

\section{The polarization of a slab of material in free space} 

Let us begin by looking at the electric dipole moment induced in an atom or molecule, $\bm p_m$, when subject to an applied electric field $\bm E$.  This is usually written as (see \cite{GriffithsEM}, Eq. 4.1).
\be
\label{eq:num_2}
\bm p_m =  \alpha_m \bm {E},
\ee
\noindent  where $\alpha_m$ is the polarizability of the atom/molecule, and is a measure of how easily the charge distribution of the molecule may be distorted by an electric field.  Next consider a slab of material made from such molecules; what happens when this slab is placed in a uniform electric field, e.g. inside a charged parallel-plate capacitor. \cite{Note1}

The dipole moment per unit volume of our material, $\bm P$, also known as the polarization, is given by $\bm P =  N \bm p_m$, where $N$ is the number of molecular dipoles per unit volume.  For a slab the field that acts to polarize the molecules is no longer just the applied field, $\bm E_0$; we also need to take into account the field at the site of any given molecule that arises from all the other charges that make up the material.  The question is thus, what is the electric field inside a neutral dielectric material in the presence of an external field? \cite{Note2}

\begin{figure}[htbp] 
\begin{center}
\mbox{\scalebox{0.25}{%
\includegraphics{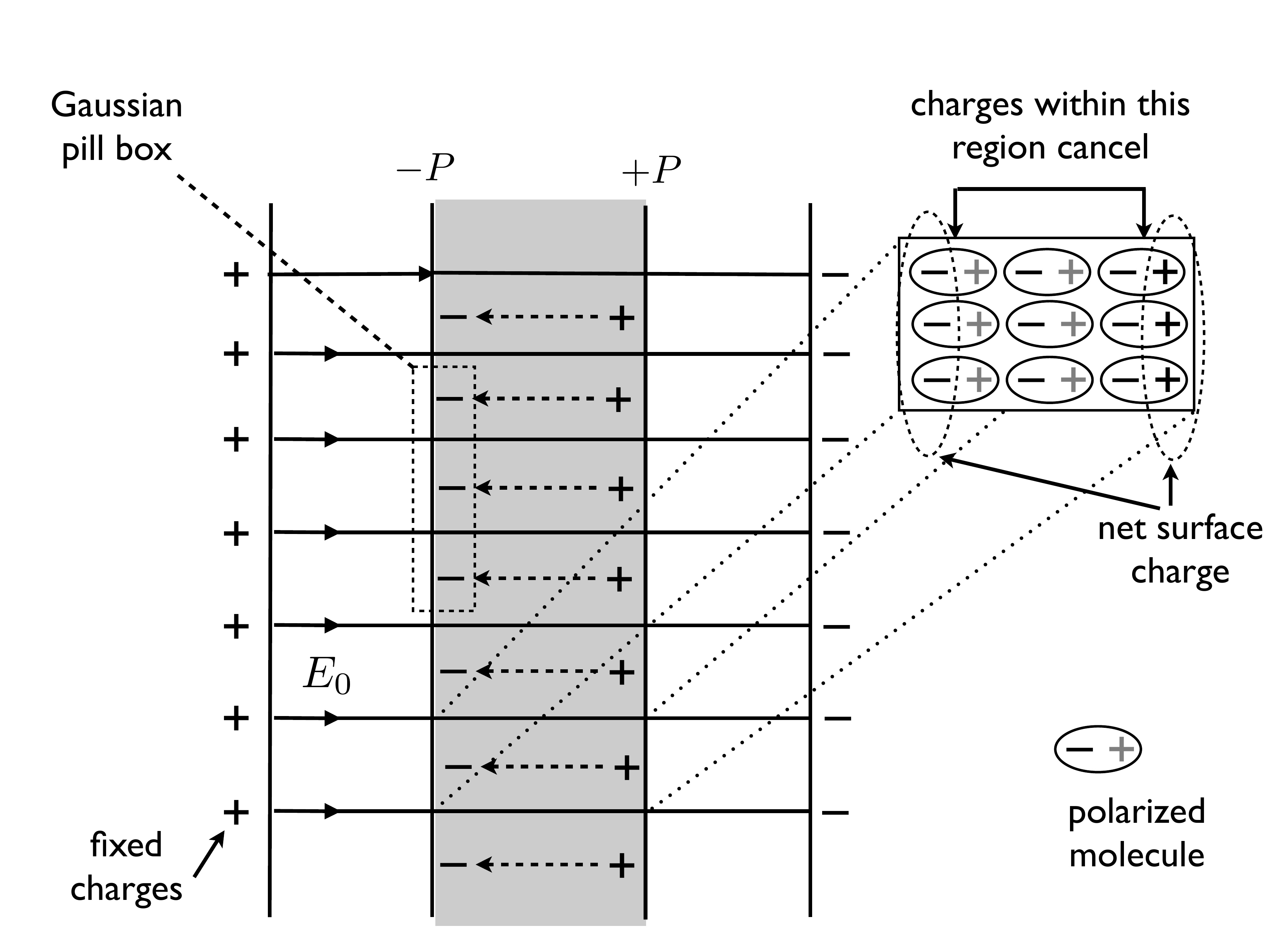}}}
\caption{A slab of a polarizable dielectric material (shaded region) placed in a uniform applied electric field $\bm E_0$ (solid arrows).  The induced surface charges on the dielectric lead to an additional electric field in the material, $\bm E_{pol}$, as indicated by the dotted arrows.  The infinite parallel-plate nature of the system means that, as with the applied field  ($\bm E_0$), the induced (depolarising) field $\bm E_{pol}$ only exists between the charged surfaces that produce it.  Inside the dielectric the net field is the sum of the applied field $\bm E_0$ and the depolarising field $\bm E_{pol}$. Also shown is a Guassian pill-box, used to evaluate the field produced by the surface charges.  The inset shows how the charges associated with the molecular polarization cancel in the bulk leaving a charge density of $\pm~P$ per unit area at the surfaces.}
\label{fig:Figure_3}
\end{center}
\end{figure}

The surface charges lead to a field inside the dielectric, $\bm E_{pol}$, often known as the depolarization field (depolarization because, in the static limit, the field produced by the displaced charges acts to counter the applied field).  The net field inside the slab, $\bm E_{net}$, is the sum of the applied field and the depolarization field, i.e.,

\be
\label{eq:num_4}
\bm E_{net}=  \bm E_0 + \bm E_{pol}.
\ee

\noindent To find $\bm E_{pol}$ we take a Gaussian pill-box that spans the interface between the dielectric slab and free space (Fig. \ref {fig:Figure_3}), and apply Gauss' law,

\be
\label{eq:num_5}
\oint\limits_S{\bm E.\bm {dA}} =  \frac {Q}{\varepsilon_0}.
\ee

\noindent The field in the integral is the net field, however the applied field makes no contribution to the integral since the charges that produce the applied field are not contained within the pill-box volume, thus only the depolarization field contributes to the integral.  If we consider a unit area of the surface then the charge enclosed in this area, $Q$, is given by $\bm P$, (the units of $\bm P$ are charge per unit area, see ~\cite{GriffithsEM} sec 4.2.2, so that numerically $P$ is equivalent to $S$).  Further, if we take the limit that the sides of the pill-box normal to the surface of the dielectric have zero extent then equation \eqref{eq:num_5} becomes,

\be
\label{eq:num_6}
\bm{E_{pol}}=  - \frac {\bm{P}}{\varepsilon_0},
\ee

\noindent where the minus sign indicates that the induced polarization is in the opposite direction to the applied field (see Fig. \ref {fig:Figure_3}).  The net field in the slab is then found from equations \eqref{eq:num_4} and \eqref{eq:num_6} to give,
\be
\label{eq:num_7}
\bm E_{net}= \bm E_0 - \frac {\bm P}{\varepsilon_0}.
\ee

We want to find the polarization in terms of the applied field $\bm E_0$, but \eqref{eq:num_7} also involves the net field in the particle. To write the net field in terms of the applied field we consider the susceptibility of the material, $\chi$, which is another measure of the ease with which the material may be polarized.  The susceptibility links the polarization $\bm P$ and the net field in the material, $\bm E_{net}$, through (see ~\cite{GriffithsEM} equation 4.30 and ~\cite{BandH} equation 2.9),
\be
\label{eq:num_8}
\bm P=\varepsilon_0 \chi \bm E_{net}.
\ee
\noindent Notice that the permittivity of free space, $ \varepsilon_0$, appears in equation \eqref{eq:num_8} as a scaling factor, ensuring that the susceptibility is dimensionless (in SI units).  To make the link between the polarization and the applied field rather than between the polarization and the net field, we need to re-express the net field of equation \eqref{eq:num_8} in terms of the applied field.  We can use \eqref{eq:num_8} to solve \eqref{eq:num_7} for $E_{net}$ in terms of $E_0$, we find,
\be
\label{eq:num_9}
E_{net}=\frac{E_0}{(1+\chi)}.
\ee
\noindent (Note that vector notation has now been dropped, in the slab geometry all fields lie on the same axis, perpendicular to the material surfaces, and the direction of the field due to the polarization is accounted for through the negative sign in equation \eqref{eq:num_6}).  Equation  \eqref{eq:num_9} shows that the strength of the field inside the slab of material is reduced by a factor of $1+\chi$ when compared to that outside.  With $\varepsilon=1+\chi$, then,
\be
\label{eq:num_10}
E_{net}=\frac{E_0}{\varepsilon}.
\ee
\noindent The quantity $\varepsilon$ is the \textit{relative} permittivity of the slab.  In terms of the relative permittivity, the polarization (see equation \eqref{eq:num_8}) can be written,
 \be
\label{eq:num_11}
P=\varepsilon_0 (\varepsilon -1)E_{net}.
\ee

\noindent We can now write the polarization in terms of the applied field, i.e. the field in the surrounding medium $E_0$, by substituting \eqref{eq:num_10} into \eqref{eq:num_11},
 \be
\label{eq:num_12}
P=\varepsilon_0 (\varepsilon -1) \frac{E_0}{\varepsilon}.
\ee

\begin{figure}[htbp] 
\begin{center}
\mbox{\scalebox{0.35}{%
\includegraphics{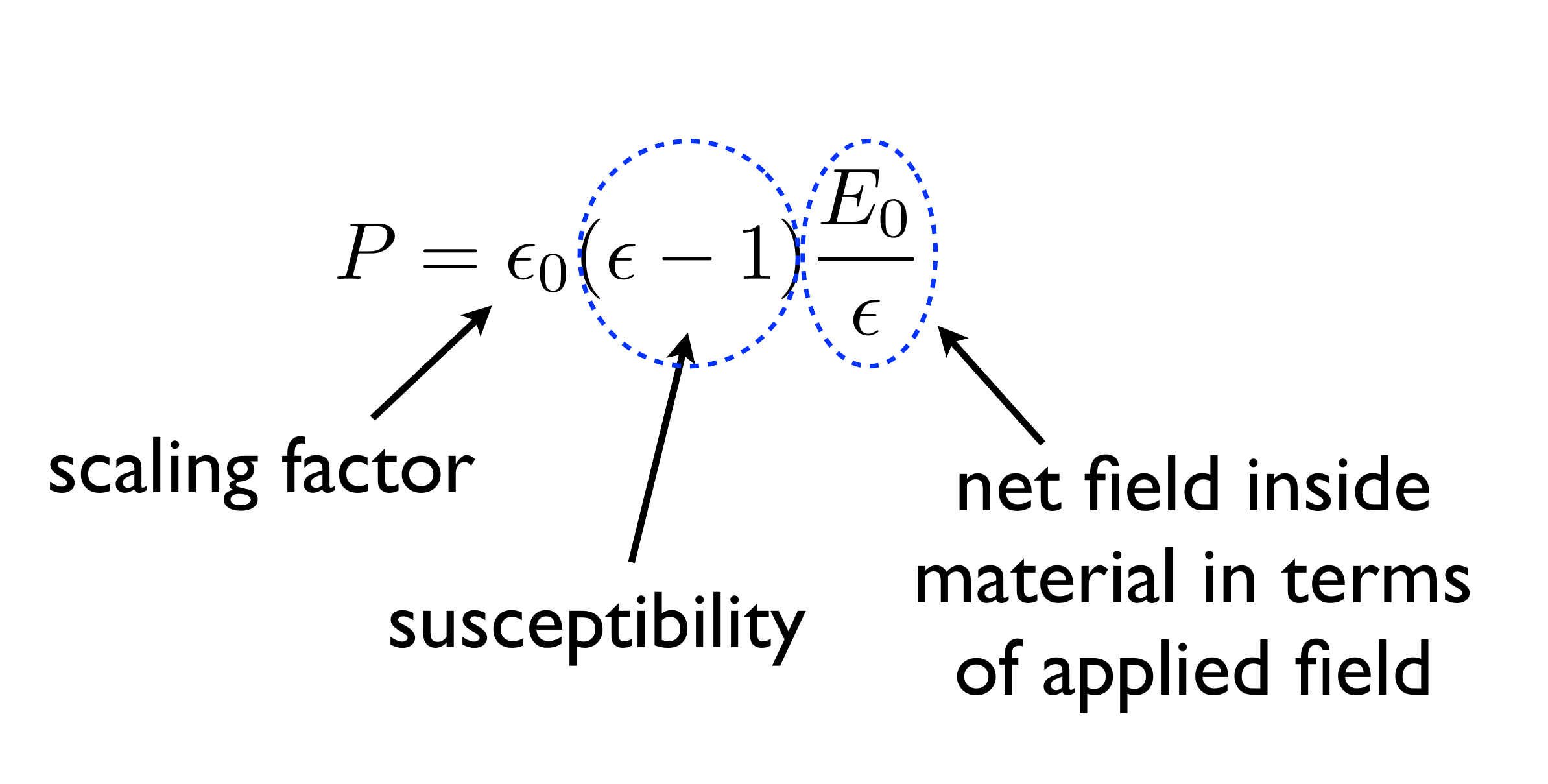}}}
\caption{The different terms that make up the polarization of a slab of dielectric material, of relative permittivity $\varepsilon$, surrounded by free space. It is the net field in the particle, ${E_0}/{\varepsilon}$, that diverges on resonance.}
\label{fig:Expression01}
\end{center}
\end{figure}

\noindent By comparing \eqref{eq:num_12} with \eqref{eq:num_8} we can see that the terms comprising the polarization are as shown in Fig. \ref {fig:Expression01}.

If we define the polarizability of the slab in a manner similar to that of an atom, see equation \eqref{eq:num_2}, then,
\be
\label{eq:num_13}
P=\alpha_V E_0,
\ee
\noindent where $\alpha_V$ is the polarizability per unit volume. By comparing equations \eqref{eq:num_12} and \eqref{eq:num_13} we find,
\be
\label{eq:num_14}
\alpha_V =\varepsilon_0 \frac{(\varepsilon -1)}{\varepsilon}.
\ee

\noindent The $(\varepsilon-1)$ term is the difference in susceptibility between the slab and its surroundings whilst $1/\varepsilon$ is there to relate the strength of the electric field inside the slab to the field outside.  There is a resonance in this polarizability when $\varepsilon=0$, this is the bulk plasmon resonance, a longitudinal mode that can not couple to light. The absence of a means to couple to light can also be seen by noting that there is no field associated with the surface polarization outside the slab (see Fig. \ref {fig:Figure_3}); however such modes can be observed using electron energy-loss spectroscopy, \cite{Powell_PR_1959_115_869} a technique that has seen a recent upsurge in interest in plasmonics with the advent of much improved technology. \cite{Scholl_Nature_2012_458_421,Schmidt_NatCom_2014_5_3604,Zhao_NL_2015_5}

\section{The polarizability of a slab of one material embedded in a slab of a different material} 

At this point we could proceed to look at the polarizability of a particle rather than a slab.  However, although there are some situations where a metallic nanoparticle may be suspended in free space, e.g. in an optical trap, \cite{Dienerowitz_JNanoPhot_2008_2_021875} in general the particle will be embedded in some dielectric medium.  We thus have to deal with (i) a change from slab geometry to particle geometry and (ii) embedding the object in a dielectric other than vacuum.  Let us examine the problem of embedding before considering the problem of geometry.

\begin{figure}[htbp] 
\begin{center}
\mbox{\scalebox{0.25}{%
\includegraphics{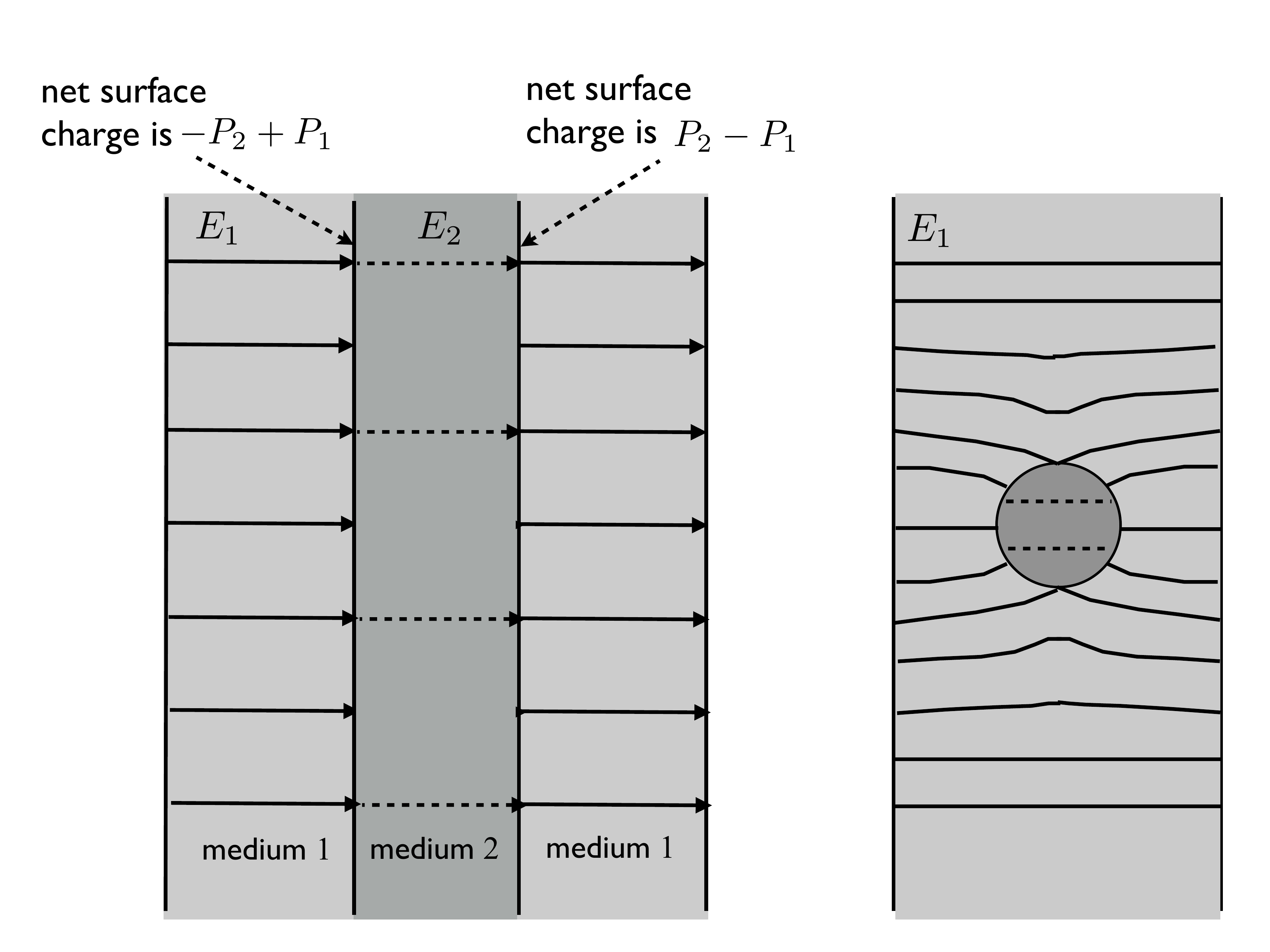}}}
\caption{Left: A slab of a polarizable dielectric (medium 2) embedded in a different dielectric (medium 1), both are immersed in the same uniform applied electric field $E_0$ (not shown), and the net field in medium 1 is $E_1$.  Note that in medium 2 the net field, $E_2$, is shown (dashed lines), this is in contrast to Fig.~\ref {fig:Figure_3} where the applied field and the depolarising field are both indicated. Right: A sphere of a polarizable dielectric (darker region) embedded in a different dielectric (lighter region), both are immersed in the same uniform applied electric field $E_0$ (not shown), and the net field in medium 1 is $E_1$.  Note that, as for the slab, in medium 2 the net field, $E_2$, is shown (dashed lines). On resonance the particle presents a cross-section to the applied field that is greater than its geometric cross-sction. \cite{Bohren_AmJPhys_1983_51_323}}
\label{fig:Figure_4}
\end{center}
\end{figure}

We are interested in the polarization of the object (medium 2) relative to its surroundings (medium 1), see Fig. \ref {fig:Figure_4} (left).  For the slab in free space we found the net field in the slab to be given by equation \eqref{eq:num_7}. For the embedded slab it is the embedding material (medium 1) rather than vacuum that is our reference material since it is the difference in material properties between the object and its immediate surroundings that gives rise to the scattering etc. of light.  As a result the field $E_0$ in \eqref{eq:num_7} needs to be replaced with $E_1$, i.e. we need to use as reference the field in the embedding medium.  Further, at the interface between the two media there will be surface charge contributions from both materials leading to a net surface charge (polarization difference) $P_{net}=P_2-P_1$.  In addition, this surface charge, when referenced to medium 1, will be screened by the material of medium 1 so that the effective surface charge will be $P_{net}/\varepsilon_1$ (see ~\cite{GriffithsEM} section 4.4.1).  The expression for the net field in the inner slab (medium 2) with reference to the embedding material (medium 1) is thus,

\be
\label{eq:num_16}
E_{net}=E_1-\frac{P_{net}}{\varepsilon_0\varepsilon_1}.
\ee

\noindent Next, following equation \eqref{eq:num_8} let us write,
\be
\label{eq:num_17}
P_{net} = \varepsilon_0 \chi_{net} E_{net},
\ee
\noindent where the net susceptibility, $\chi_{net}$, is the difference in susceptibility of the two materials, i.e. $\chi_{net}=\chi_2-\chi_1$.  Substituting \eqref{eq:num_16} into \eqref{eq:num_17} and solving for $P_{net}$ gives,

\be
\label{eq:num_19}
P_{net}=\varepsilon_0\frac{\chi_{net}}{1+\chi_{net}/\varepsilon_1}E_1.
\ee

\noindent If we now make use of the fact that $(1+\chi_1)=\varepsilon_1$ and $(1+\chi_2)=\varepsilon_2$, then, $\chi_{net}=\chi_2-\chi_1=\varepsilon_2-\varepsilon_1$, so that \eqref{eq:num_19} becomes,

\be
\label{eq:num_21}
P_{net}=\varepsilon_0 (\varepsilon_2-\varepsilon_1) \frac{\varepsilon_1E_1}{\varepsilon_2}.
\ee

\begin{figure}[htbp] 
\begin{center}
\mbox{\scalebox{0.3}{%
\includegraphics{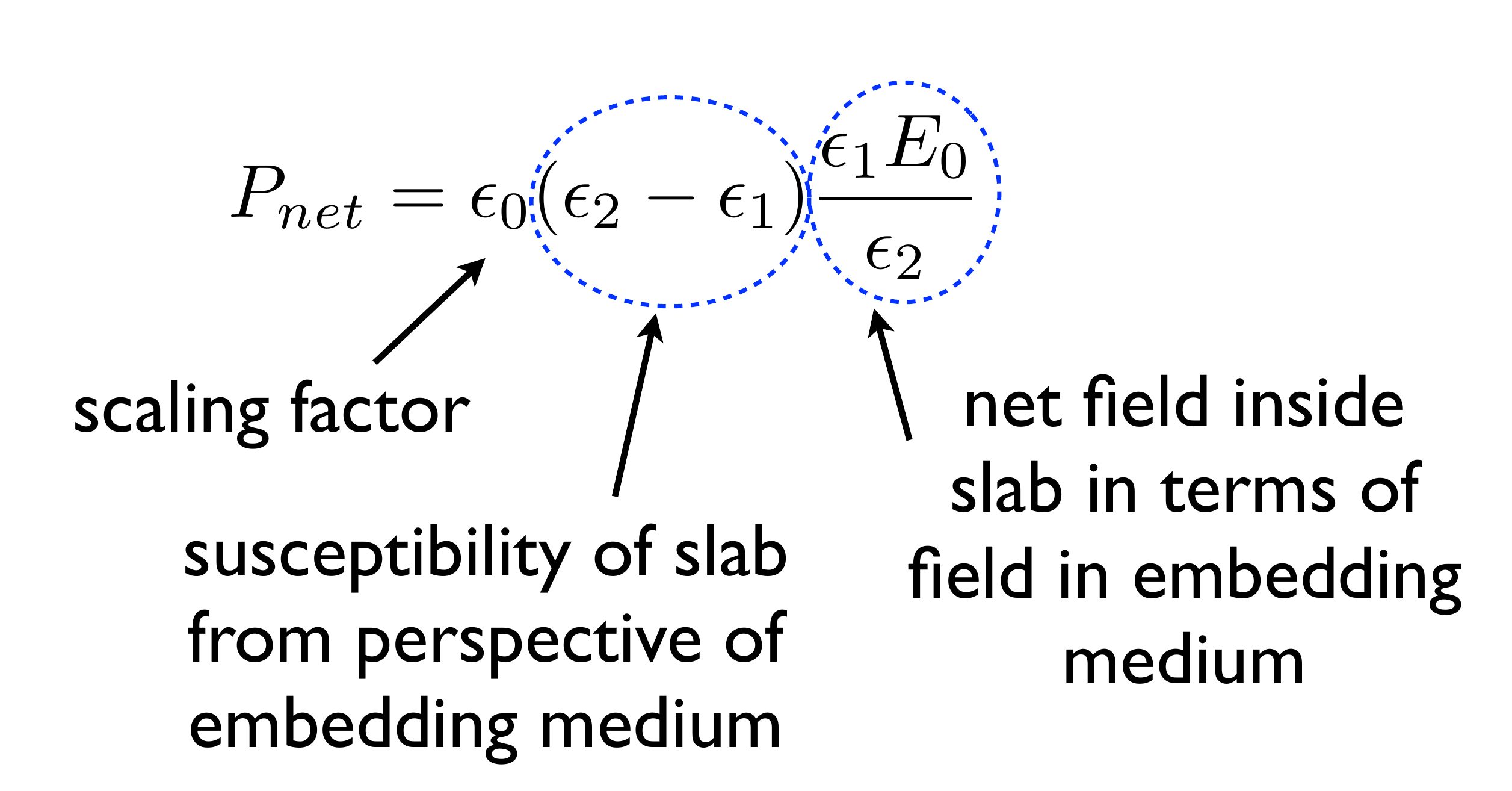}}}
\caption{The different terms that make up the polarization of a slab of one material (medium 2) embedded in a different material (medium 1). Note that the resonance condition, for which the field in the particle diverges, occurs when $\varepsilon_2=0$, it is thus independent of the embedding medium, this is because there in the slab geometry there is no field associated with the (surface) polarization in the embedding medium.}
\label{fig:Expression02}
\end{center}
\end{figure}

\noindent If we compare \eqref{eq:num_21} with \eqref{eq:num_12} we see that if they are to take the same basic form then the final term in equation \eqref{eq:num_21}, $\varepsilon_1E_1/\varepsilon_2$, should be equal to the net field in the material, $E_2$, written in terms of the applied field which, in this case, is $E_1$, the field in the embedding medium.   Is $\varepsilon_1E_1/\varepsilon_2$ equal to $E_2$? Consider each slab on its own, embedded in vacuum, as we did in section II.  Then by analogy with \eqref{eq:num_10} we can write $E_1=E_0/\varepsilon_1$ and $E_2=E_0/\varepsilon_2$.  These relationships will still apply in the present case since, if we butt one slab up against another, no new effects are produced; all the (surface) charges involved are bound charges, we thus have, $\varepsilon_1 E_1=\varepsilon_2E_2$. We can now see that the term at the end of \eqref{eq:num_21}, ${\varepsilon_1E_1}/{\varepsilon_2}$ is indeed the field in medium 2, i.e. $E_2$. The terms comprising the polarization of the embedded slab are shown in figure \ref {fig:Expression02}.

Finally let us work out the polarizability (per unit volume).  For a slab in free space the polarizability is defined as $P=\alpha_V E_0$, see equation \eqref{eq:num_13}.  For the slab embedded in a dielectric of relative permittivity $\varepsilon_1$ we can write,
\be
\label{eq:num_23}
P_{net}=\alpha_V E_1.
\ee
\noindent Comparing \eqref{eq:num_23} and \eqref{eq:num_21} we find that the polarizability per unit volume of the embedded slab is thus,
\be
\label{eq:num_24}
\alpha_V=\varepsilon_0\varepsilon_1 \frac{(\varepsilon_2-\varepsilon_1)}{\varepsilon_2}.
\ee
\noindent As with the slab in free space there is a resonance if $\varepsilon_2=0$.  The term $(\varepsilon_2-\varepsilon_1)$ comes from the difference in susceptibility of the two materials whilst the $\varepsilon_2$ in the denominator relates the field inside the slab to that outside; $\varepsilon_0$ is a scaling factor and the remaining $\varepsilon_1$ arises from our choice of reference material, i.e. we are working with reference to medium 1. 

Equation \eqref{eq:num_21} can be obtained from \eqref{eq:num_12} by (i) replacing $\varepsilon$ (perhaps more properly $\varepsilon/1$) with $\varepsilon_2/\varepsilon_1$, and (ii) ensuring that the applied field is appropriate to the situation, i.e. $E_0$ when free space is the bounding medium and $E_1$ when medium 1 is the bounding medium.  If this prescription is followed then expressions \eqref{eq:num_12} and \eqref{eq:num_21} are the same (see ~\cite{Jones_PR_1945_68_93} and section on p 41 of ~\cite{LLandP}).

\section{The polarizability of small sphere of one material embedded in a different material} 

We now look at a small (relative to the wavelength of interest) sphere of the material and the polarization it acquires in the presence of a uniform applied electric field, see Fig. \ref {fig:Figure_4} (right) above. As for the dielectric slab, we assume the field inside the sphere is homogeneous. The change from a planar geometry leads to a modification of the surface charge distribution around the sphere which in turn leads to a change in the field produced inside the sphere.  The change in geometry also means that these surface charges will produce a field outside the sphere, thereby providing a means for external fields to couple to the polarization of the sphere.

The last term in  equation \eqref{eq:num_16}, $P_{net}/\varepsilon_0\varepsilon_1$, is the field that arises in the embedded slab due the the surface charges $(P_{net})$ as seen from the perspective of medium 1.  In the case of our small sphere of material, this field will be different from the slab by some multiplicative factor $L$ (the shape factor), the depolarizing field is thus, \cite{Note3}
\be
\label{eq:num_25}
E_{pol}=-L\frac{P_{net}}{\varepsilon_1\varepsilon_0},
\ee
\noindent We can now modify the equation for the net field \eqref{eq:num_10} to give the net field in the sphere, with reference to the embedding medium (medium 1), as,
\be
\label{eq:num_26}
E_{net}=E_1-L\frac{P_{net}}{\varepsilon_0\varepsilon_1}.
\ee
\noindent If we use equation, \eqref{eq:num_26}, instead of \eqref{eq:num_16} and proceed with the analysis that follows \eqref{eq:num_16} then we find that the polarization of the sphere is given by,
\be
\label{eq:num_27}
P_{net}=\varepsilon_0 (\varepsilon_2-\varepsilon_1) \frac{\varepsilon_1E_1}{\varepsilon_1+L(\varepsilon_2-\varepsilon_1)}.
\ee

\begin{figure}[htbp] 
\begin{center}
\mbox{\scalebox{0.3}{%
\includegraphics{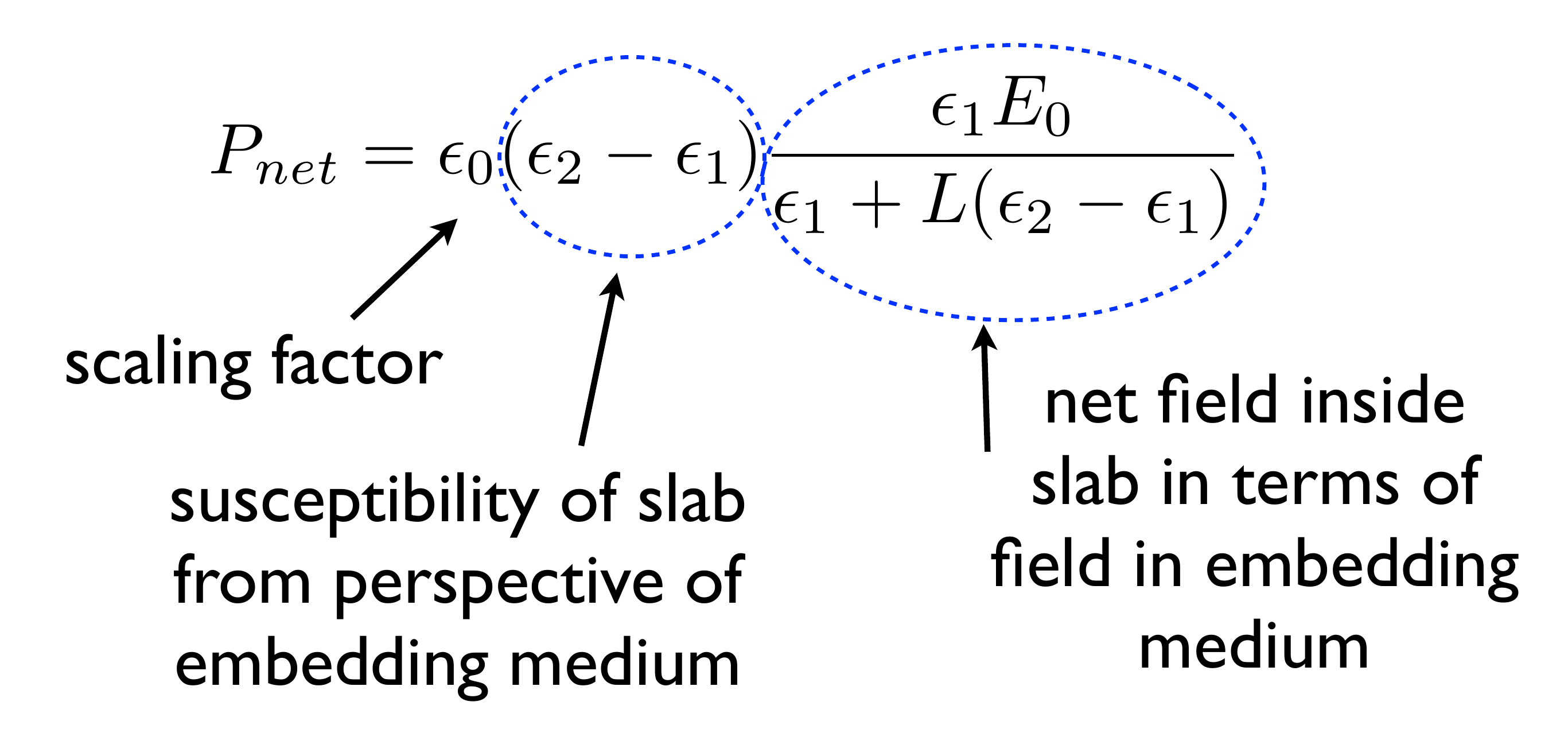}}}
\caption{The different terms that make up the polarization of a sphere (or other shape) of one material (medium 2) embedded in a different material (medium 1).}
\label{fig:Expression03}
\end{center}
\end{figure}

\noindent By comparing \eqref{eq:num_27} and \eqref{eq:num_21} we can see that the different terms in this expression for the polarization are the same as those for the embedded slab, with the exception of the field inside the object, see figure \ref {fig:Expression03}. 

\noindent The polarizability per unit volume of the sphere can now be found with the aid of \eqref{eq:num_23}, it is,
\be
\label{eq:num_28}
\alpha_V=\frac{\varepsilon_0\varepsilon_1(\varepsilon_2-\varepsilon_1)}{(\varepsilon_1+L(\varepsilon_2-\varepsilon_1))},
\ee
\noindent so that the polarizability is,
 \be
\label{eq:num_29}
\alpha=V\frac{\varepsilon_0\varepsilon_1(\varepsilon_2-\varepsilon_1)}{(\varepsilon_1+L(\varepsilon_2-\varepsilon_1))},
\ee
\noindent where V is the volume of the particle.  As before, the $(\varepsilon_2-\varepsilon_1)$ in the numerator arises from the difference in susceptibility of the two materials, whilst the denominator, $(\varepsilon_1+L(\varepsilon_2-\varepsilon_1))$, is associated with the electric field inside the particle. The resonance condition now occurs when,
\be
\label{eq:L_resonance}
\varepsilon_2=\left(\frac{-1}{L}+1\right)\varepsilon_1.
\ee

For a spherical object where $L=1/3$ then, if we assume that the embedding medium is vacuum, ($\varepsilon_1=1$), and we substitute $L=1/3$ into \eqref{eq:num_29}, and further note that to find the polarizability (rather than the polarizability per unit volume) we need to multiply by the volume of the sphere, then we recover the expression we began with, equation \eqref{eq:num_1}. Having carried out the analysis above we can see that in \eqref{eq:num_1} the $\varepsilon-1$ term arises from the difference between the susceptibility of the material from which the sphere is made and the susceptibility of its surroundings (here vacuum), whilst the $\varepsilon+2$ term arises from the way the shape of the particle, and in particular the resulting surface-charge distribution, dictates the net field inside the particle.

\begin{figure}[htbp] 
\begin{center}
\mbox{\scalebox{0.5}{%
\includegraphics{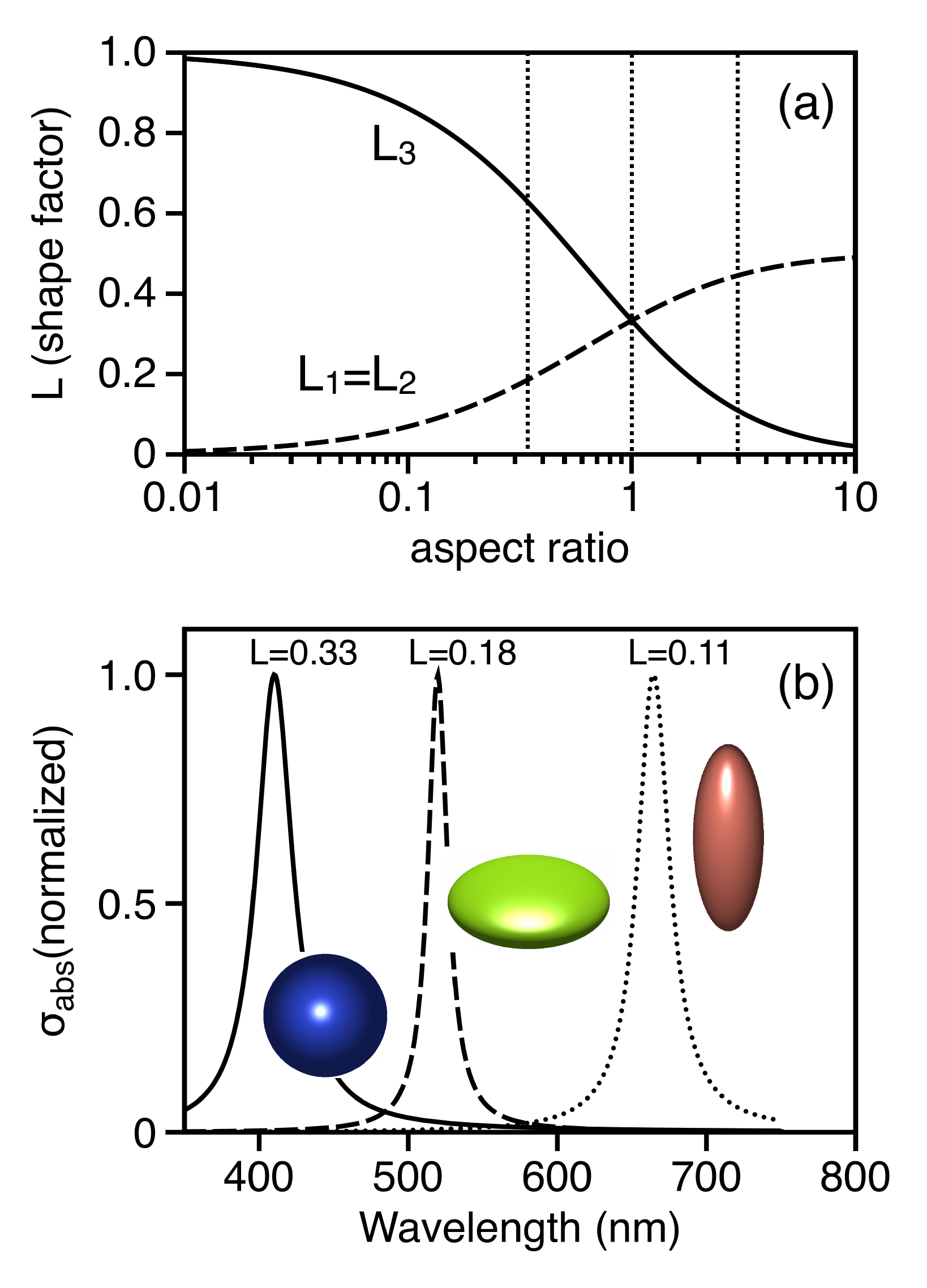}}}
\caption{Upper panel (a). The depolarization factors, $L_j$, for spheroids ($r_1=r_2\ne r_3$) as a function of the aspect ratio, $r_3/r_1$. For a sphere $r_1=r_2=r_3$, the aspect ratio is 1, and the depolarization factors are all equal to $1/3$. Lower panel (b). Absorption spectra caluclated using equations \ref{eq:num_sig_scatt} and \ref{eq:num_29} for three different spehroids: a sphere, a platelet and a rod having aspect ratios of 1, 1/3, and 3 respectively (corresponding to the three vertical dottred lines in (a)); the platelet and rod are illuminated with $E$ along a long axis. The relevant shape factors are indicated. Also shown in (b) are synthetically rendered images where the colour is based on the calculated spectrum.} 
\label{fig:Figure_5}
\end{center}
\end{figure}

Standard textbook treatments of how electromagnetic fields behave in matter have something very similar.  In particular, the Clausius-Mossotti relation provides an expression for the polarizability per molecule, $\alpha_m$, of a material, which is usually derived by finding the electric field inside a small spherical void in the material. \cite{Note4}  The similarity of the Clausius-Mossotti relation and equation~\eqref{eq:num_1} comes from the fact the the Clausius-Mossotti equation for the molecular polarizability involves the susceptibility $\chi$ $(=\varepsilon-1)$ in the numerator, and, because it uses the same spherical geometry, the factor $\varepsilon+2$ appears in the denominator (see for example \cite{PandP} and \cite{AandM}). The value of the permittivity at which resonance occurs is $\varepsilon_2=-2$, and the frequency (wavelength) at which this occurs will in turn be dictated by the dispersion of the relative permittivity of the material, i.e. how the permittivity varies with frequency (wavelength), Fig. \ref {fig:Figure_2}.  For a simple free-electron metal where the frequency dependent permittivity may be described by the Drude-Lorentz formula, \cite{Fox_OPS}
\be
\label{eq:num_aa}
\varepsilon(\omega)=1-\frac{\omega_P^2}{\omega^2},
\ee
\noindent so that the frequency at which $\varepsilon_2=-2$ occurs is $\omega=\omega_P/\sqrt{3}$. The effect of shape on the  resonance frequency is shown in Fig. \ref{fig:Figure_5}.

Resonance in a particle is thus determined by the depolarization field, which in turn depends on the shape of the particle; it is for this reason that shape is such a powerful controlling factor in determining resonance frequencies in plasmonics.  

At this point it is worth taking some input from the fuller solution to the problem. Finding the polarization of a small ellipsoidal particle \cite{Note5} in the electrostatic limit is usually accomplished by solving Laplace's equation for the electric potential, making use of the boundary conditions at the material interfaces. Working out the value of $L$ for particles of different shape is easily done.\cite{NandH}  For a general ellipsoid the shape factor depends on the symmetry axis considered.  There are three shape factors, one for each symmetry axis $j$, the factors $L_j$ satisfying the sum rule $\sum_j L_j=1$, the $L_j$ are given by,~\cite{BandH, Myroshnychenko_ChemSocRev_2008_37_1792}

\be
\label{eq:L_j}
L_j=\frac{r_1r_2r_3}{2} \int_0^\infty \frac{ds}{\left(s+r_j^2\right)\sqrt{\left(s+r_1^2\right)\left(s+r_2^2\right)\left(s+r_3^2\right)}},
\ee

\noindent where $r_j$ are the radii of the ellipsoid, see Fig. \ref{fig:Figure_5}.

For rod shaped particles ($r_3>r_1,r_2$) excited by an electric field along their long axis, the relevant depolarization factor is $L_3$ and is smaller than that of a sphere. From equation (\ref{eq:L_resonance}) we see that resonance for such a particle requires a more negative permittivity which, looking at Fig. \ref{fig:Figure_2}, will occur at longer wavelengths - the change in the shape results in the resonance being redshifted. Another common particle geometry is the disc or platelet ($r_3<r_1,r_2$). When the applied electric field is parallel to the short axis the relevant shape axis is again $L_3$ but now it is larger than 1/3 and the resonance is blue shifted. However, for discs and platelets the axis of interest is usually across the disc/platelet. In this case the appropriate $L_j$ is $L_1 (=L_2)$ so that the resonance will again be redshifted. The extent of the changes in the resonance condition arising from different particle shapes are illustrated in Fig. \ref{fig:Figure_5}.

So far we have ignored damping, however the permittivity of the metal is in general a complex quantity owing to damping of the free-electron response. Including this complex nature leads to the following expression for the permittivity,~\cite{Fox_OPS}
\be
\label{eq:num_bb}
\varepsilon(\omega)=1-\frac{\omega_P^2}{\omega^2+i\gamma \omega}=1-\frac{i\sigma(\omega)}{\varepsilon_0\omega},
\ee
\noindent where $\gamma$ is the scattering rate. (For comparison the permittivity is also been given in terms of the complex conductivity, $\sigma(\omega)$). The complex nature of the permittivity has two consequences.  First, the resonance condition can not be perfectly met.\cite{Note6} Second, the resonance has a finite amplitude and width. 

If the embedding medium has a permittivity of $\varepsilon_1$, then the resonance condition for the sphere becomes $\varepsilon_2=-2\varepsilon_1$, the resonance condition is thus sensitive to the environment, an effect that is exploited in particle-plasmon based biosensors where target molecules bound to the particle change the effective permittivity of the medium surrounding the particle, thereby altering the resonance condition. \cite{Whitney_JPCB_2005_109_20522, Raschke_NL_2003_3_935, Murray_NL_2006_6_1772}

Finally, we should consider some of the limitations of the quasistatic approach. For gold and silver spheres, the quasistatic approximation for the polarizability seems to hold reasonably well up to a diameter of $\sim$ 50 nm, \cite{Kelly_JPCB_2003_107_668} for other shapes the size range over which the quasistatic approach holds is more restrictive. Modelling the response of larger particles requires radiative damping and dynamic depolarization to both be taken into account (dynamic depolarization refers to the fact that the depolarising field at a given time depends on the charge distributions at some earlier time, i.e. it is retarded).  This can be done in an analytic (and approximate) way using, for example, the modified long-wavelength approach~\cite{Meier_OL_1983_8_581} exploited by Kuwata et al. \cite{Kuwata_APL_2003_83_4625} and discussed in detail by Moroz. \cite{Moroz_JOSAB_2009_26_517} The most accurate method is the full electrodynamic solution to the polarizability, based on solving Maxwell's equations, thereby allowing a greater range of particle sizes to be considered, \cite{Tcherniak_NL_2010_10_1398} and effects such as higher order plasmon modes to be included. \cite{Burrows_OE_2010_18_3187}

Many interesting aspects have been left out of the discussion above: the role of substrates, \cite{Knight_NL_2009_9_2188} the role of coatings on particles (i.e. metallic cores with a dielectric coating), and proximity to adjacent particles. \cite{Rechberger_OC_2003_220_137} There is also the fascinating topic of interactions between particles in arrays. \cite{Abajo_RMP_2007_79_1267} Alternative approaches are needed for modelling these systems, such as the boundary-element method, \cite{Hohenester_CompPhysComm_2012_183_370} the finite-difference time-domain technique, \cite{Parsons_JMO_2010_57_356} and the discrete dipole-approximation. \cite{Noguez_JPCC_2007_111_3806} In addition, these alternatives also allow particles of arbitrary shape to be investigated.

Despite its many shortcomings the quasistatic model for the polarizability of metallic nanoparticles is a helpful and often appropriate model in plasmonics. Perhaps its greatest asset is that it allows a range of properties to be explored analytically. Seeing where the different terms in the polarizability come from helps to build an appreciation of the underlying physics.
 
\begin{acknowledgments}
The author is indebted to Baptiste Augui\'e, Euan Hendry, Chris Burrows and Alastair Humphrey for many thought-provoking discussions. The assistance of Alastair Humphrey, Martin Gentile and Ian Hooper with some of the figures is gratefully acknowledged. This work was sponsored both by The Leverhulme Trust and by the The Royal Society. The many helpful and constructive comments from the reviewers is also appreciated.

\end{acknowledgments}

\begin{appendices}

\section*{APPENDIX: Conventions}

\setcounter{equation}{0}
\renewcommand{\theequation}{A{\arabic{equation}}}

The expressions for the polarizability (per unit volume) derived here agree with those given by Le Ru and Etchegoin, \cite{LRandE} but are different from those given by others, for example Bohren and Huffman, \cite{BandH}  Myroshnychenko \textit{et al.} \cite{Myroshnychenko_ChemSocRev_2008_37_1792} and Kreibig and Vollmer; \cite{KandV} different authors adopt different conventions (see \cite{LRandE} p 74), specifically by adopting different definitions of the polarization.  Here we defined the polarizability (per unit volume) through the equation, \textit{net polarization = polarizability $\times$ applied field}, i.e. equation \eqref{eq:num_13}.  Le Ru and Etchegoin adopt the same convention, as do Novotny and Hecht.  Bohren and Huffman chose to define the polarizability (per unit volume) through the equation, \textit{net polarization = permittivity of embedding medium $\times$ polarizability $\times$ applied field} (their equation 5.15), consequently their polarizability (per unit volume) is, $\alpha_V=(\varepsilon_2-\varepsilon_1)/(\varepsilon_1+L(\varepsilon_2-\varepsilon_1))$.  Kreibig and Vollmer chose to define the polarizability (per unit volume) through the equation, \textit{net polarization = relative permittivity of embedding medium $\times$ polarizability $\times$ applied field} (their equation 2.12a), consequently their polarizability (per unit volume) is, $\alpha_V=\varepsilon_0(\varepsilon_2-\varepsilon_1)/(\varepsilon_1+L(\varepsilon_2-\varepsilon_1))$.  Lastly, Myroshnychenko \textit{et al.} state their polarizability (per unit volume) as, $\alpha_V=\varepsilon_1(\varepsilon_2-\varepsilon_1)/(\varepsilon_1+L(\varepsilon_2-\varepsilon_1))$, from which we infer that they have assumed the polarizability to be defined through \textit{net polarization = permittivity of free space $\times$ polarizability $\times$ applied field}. Since the different conventions adopted by various authors can easily lead to confusion it is worth looking at an example, the scattering cross-section.

Light incident on a particle creates an oscillating dipole moment, and oscillating dipoles radiate; it is this re-radiated light that we refer to as scattering.  The scattering cross-section is defined as the ratio of the power radiated by a dipole, $W$, to the intensity of light incident upon it, $I$, i.e.,
\be
\label{eq:num_32}
\sigma_{sca}=\frac{W}{I}.
\ee
The (time averaged) power radiated by a dipole is (see ~\cite{NandH} equation 8.71),
\be
W=\frac{|p|^2\omega^4n^3}{12\pi\varepsilon_0\varepsilon c^3},
\ee
\noindent where $p$ is the dipole moment of the particle, $\omega$ is the angular frequency of the incident light, $n$ is the refractive index of the medium in which the dipole (particle) is embedded, and $\varepsilon$ is the relative permittivity of the medium in which the dipole is embedded.  (The dependence of the radiated power on the refractive index of the embedding medium is discussed by Barnet \textit{et al.} \cite{Barnett_JPhysB_1996_29_3763}).  Re-writing this last equation in the notation used in the present article,
\be
\label{eq:num_33}
W=\frac{|p|^2\omega^4\sqrt{\varepsilon_1}}{12\pi\varepsilon_0 c^3}.
\ee
The (time averaged) intensity of a plane wave is given by (see ~\cite{LCandL} equation section 28.4),
\be
 \label{eq:num_34}
I=\frac{1}{2}\sqrt{\frac{\varepsilon}{\mu}}E^2.
\ee
Again, rewriting this in the notation used in the present article and, providing we assume $\mu=\mu_0$ (i.e. the relative permeability of the embedding medium is unity), then,
 \be
 \label{eq:num_35}
I=\frac{1}{2}\sqrt{\varepsilon_1}\varepsilon_0 c E^2.
\ee
Substituting  \eqref{eq:num_33} and  \eqref{eq:num_35} into  \eqref{eq:num_32} we obtain,
\be
\label{eq:num_36}
\sigma_{sca}=\frac{k^4_0 |p|^2}{6\pi\varepsilon_0^2 E^2}.
\ee
Now recall that for a particle the dipole moment $p$ is just the polarization $P$ per unit volume of the particle multiplied by the volume, i.e. $p=VP$, and that from \eqref{eq:num_13} we have that $P=\alpha_V E$ with $\alpha_V$ the polarizability per unit volume.  The polarizability is then $\alpha=V\alpha_V$, so that $p=VP=\alpha E$.  Substituting $p=\alpha E$ into \eqref{eq:num_36} we recover equation \eqref{eq:num_sig_scatt},
\be
\tag{\ref{eq:num_sig_scatt}}
\sigma_{sca}=\frac{k^4_0 |\alpha|^2}{6\pi\varepsilon_0^2}.
\ee
The expression given by Bohren and Huffman (see ~\cite{BandH} equation 5.19) is,
\be
\label{eq:num_38}
\sigma_{sca}=\frac{k^4_M |\alpha|^2}{6\pi}.
\ee
where $k_M$ is the wavevector in the embedding medium ($k_M=k_0\sqrt{\varepsilon_M}$).  Expressions \eqref{eq:num_sig_scatt} and \eqref{eq:num_38} give exactly the same result, provided the polarizability appropriate to the definition of the polarization in each case is used. Notice that the equation for the scattering cross-section contains $k^4=\omega^4/c^4$, i.e. the strength of the scattering depends on the 4th power of the frequency. This is Rayleigh scattering and accounts for the variation with wavelength, for example that of the scattering from a $\mathrm{TiO_2}$ nanosphere shown in Fig. 1. The resonance feature for the gold sphere in Fig. 1 arises from the (plasmon) resonance in the polarizability, superimposed on a background of Rayleigh scattering.

\end{appendices}





\begin{thebibliography}{99}

\bibitem{Barber_Archaeometry_1990_32_33}
D.~J. Barber and I.~C. Freestone, ``An investigation of the origin of the colour of the lycurgus cup by analytical transmission electron microscopy"
\newblock Archaeometry {\bf 32}, 33-45 (1990).

\bibitem{Faraday_PhilTrans_1857_147_145}
M.~Faraday, ``Experimental Relations of Gold (and other Metals) to Light"
\newblock Phil. Trans. Royal Soc. {\bf 147}, 145-181
  (1857).

\bibitem{Mie_AnnPhys_1908_25_377}
G.~Mie, ``Beitr\"age zur Optik tr\"uber Medien, speziell kolloidaler Metall\"osungen"
\newblock Ann. d. Phys. {\bf 25}, 377-445 (1908).

\bibitem{Gramotnev_NatPhot_2014_8_13}
D.~K. Gramotnev and S.~I. Bozhevolnyi, ``Nanofocusing of electromagnetic radiation"
\newblock Nature Phot. {\bf 8}, 13-22 (2013).

\bibitem{Murray_NL_2006_6_1772}
W.~A. Murray, J.~R. Suckling, and W.~L. Barnes, ``Overlayers on silver nanotriangles: Field confinement and spectral position of localized surface plasmon resonances"
\newblock Nano Lett. {\bf 6}, 1772-1777 (2006).

\bibitem{LRandE}
E.~C. {Le Ru} and P.~G. Etchegoin,
\newblock {\em {Principles of Surface-Enhanced Raman Spectroscopy and related
  plasmonic effects}},
\newblock (Elsevier, 1st edition, 2009).

\bibitem{Andrew_JMO_1997_44_395}
P.~Andrew, S.~C. Kitson, and W.~L. Barnes, ``Surface-plasmon energy gaps and photoabsorption"
\newblock J, Mod. Opt. {\bf 44}, 395-406 (1997).

\bibitem{Gruhlke_PRL_1986_56_2838}
R.~Gruhlke, W.~Holland, and D.~Hall, ``Surface plasmon cross coupling in molecular fluorescence near a corrugated thin metal film"
\newblock Phys. Rev. Lett. {\bf 56}, 2838-2841 (1986).

\bibitem{Kitson_PRB_1995_52_11441}
S.~C. Kitson, W.~L. Barnes, and J.~R. Sambles, ``Surface-plasmon energy gaps and photoluminescence"
\newblock Phys. Rev. B. {\bf 52}, 11441-11446 (1995).

\bibitem{Torma_RepProgPhys_2015_78_013901}
P.~T{\"{o}}rm{\"{a}} and W.~L. Barnes, ``Strong coupling between surface plasmon polaritons and emitters: a review"
\newblock Rep. Prog. Phys. {\bf 78}, 013901-1--34 (2015).

\bibitem{Polyushkin_NL_2011_11_4718}
D.~K. Polyushkin, E.~Hendry, E.~K. Stone, and W.~L. Barnes, ``THz generation from plasmonic nanoparticle arrays"
\newblock Nano Lett. {\bf 11}, 4718-4724 (2011).

\bibitem{Fischer_PRL_1989_62_458}
U.~C. Fischer and D.~W. Pohl, ``Observation of single-particle plasmons by near-field microscopy"
\newblock Phys. Rev. Lett. {\bf 62}, 458-462 (1989).

\bibitem{Atwater_NatMat_2010_9_205}
H.~A. Atwater and A.~Polman, ``Plasmonics for improved photovoltaic devices"
\newblock Nature Mat. {\bf 9}, 205-213 (2010).

\bibitem{Loo_NL_2005_5_709}
C.~Loo, A.~Lowery, N.~J. Halas, J.~West, and R.~Drezek, ``Immunotargeted nanoshells for integrated cancer imaging and therapy"
\newblock Nano Lett. {\bf 5}, 709-711 (2005).

\bibitem{Willets_AnnuRevPhysChem_2007_58_267}
K.~A. Willets and R.~P. {Van Duyne}, ``Localized surface plasmon resonance spectroscopy and sensing"
\newblock Ann. Rev. Phys. Chem. {\bf 58}, 267-297 (2007).

\bibitem{Wu_AdvFuncMat_2015}
M. Wu, B. Ma, T. Pan, S. Chen and J. Sun, ``Silver-nanoparticle-colored cotton fabrics with tunable colors and durable antibacterial and self-healing superhydrophobic properties"
\newblock Adv. Func. Mat. {\bf 26}, 569-576 (2016)

\bibitem{Meinzer_NatPhot_2014_8_889}
N.~Meinzer, W.~L. Barnes, and I.~R. Hooper, ``Plasmonic meta-atoms and metasurfaces"
\newblock Nature Phot. {\bf 8}, 889-898 (2014).

\bibitem{LandH}
D.~W. Lynch and W.~R. Hunter,
\newblock {Comments on the Optical Constants of Metals and an Introduction to the Data of Several Metals},
\newblock in {\em Handbook of Optical Constants of Solids}, edited by E.~D.
  Palik, (Academic Press Inc., 1985.), p. 275-367

\bibitem{BandH}
C.~F. Bohren and D.~R. Huffman,
\newblock {\em {Absorption and scattering of light by small particles}},
\newblock (Wiley-VCH, 2004.)

\bibitem{Jones_PR_1945_68_93}
R.~C. Jones, ``A generalization of the dielectric ellipsoid problem"
\newblock Phys. Rev. {\bf 68}, 93-96 (1945).

\bibitem{GriffithsEM}
D.~J. Griffiths,
\newblock {\em {Introduction to Electrodynamics}},
\newblock (Prentice-Hall, 3rd edition, 1999.)

\bibitem{Luther_NatMat_2011_10_361}
J.~M. Luther, P.~K. Jain, T.~Ewers, and A.~P. Alivisatos, ``Localized surface plasmon resonances arising from free carriers in doped quantum dots"
\newblock Nature Mat. {\bf 10}, 361-366 (2011).

\bibitem{Note1}
We assume the charges on the plates of the capacitor are fixed so that the
  field they produce between the plates, $\protect \bm {E}_0$, is both
  homogeneous and constant.

\bibitem{Note2}
In the context of the present discussion, metals can be considered as a subset
  of dielectrics in the sense that their permittivity is not infinite (we are
  not interested here in perfect metals (perfect conductors) where the electric
  field is totally excluded from the metal), there will thus be an electric
  field present inside the metal.

\bibitem{Powell_PR_1959_115_869}
C.~J. Powell and J.~B. Swan, ``Origin of the characteristic electron energy losses in aluminum"
\newblock Phys. Rev. {\bf 115}, 869-875 (1959).

\bibitem{Scholl_Nature_2012_458_421}
J.~A. Scholl, A.~L. Koh, and J.~A. Dionne, ``Quantum plasmon resonances of individual metallic nanoparticles"
\newblock Nature {\bf 483}, 421-427 (2012).

\bibitem{Schmidt_NatCom_2014_5_3604}
F.-P. Schmidt et~al., ``Universal dispersion of surface plasmons in flat nanostructures"
\newblock Nature Comm. {\bf 5}, 3604-1--6 (2014).

\bibitem{Zhao_NL_2015_5}
M.~Zhao et~al., ``Visible Surface Plasmon Modes in Single Bi$_2$Te$_3$ Nanoplate"
\newblock Nano Lett. {\bf 15}, 2-6 (2015).

\bibitem{Dienerowitz_JNanoPhot_2008_2_021875}
M.~Dienerowitz, M.~Mazilu, and K.~Dholakia, ``Optical manipulation of nanoparticles: a review"
\newblock J. Nanophot. {\bf 2}, 021875-1--32 (2008).

\bibitem{Bohren_AmJPhys_1983_51_323}
C.~F. Bohren, ``How can a particle absorb more than the light incident on it?"
\newblock Am. J. Phys. {\bf 51}, 323-327 (1983).

\bibitem{LLandP}
E.~M. Landau, L.~D. Lifshitz, and L.~P. Pitaevskii,
\newblock {\em {Electrodynamics of continuous media, Vol 8 of Course of
  Theoretical Physics}},
\newblock (Elsevier, 2nd edition, 2006).

\bibitem{Note3}
We retain $L$ here rather than insert the numerical value, so as to allow the
  derivation of a general expression.

\bibitem{Note4}
To understand the modes supported by voids in metals we need to swap the roles
  of $\varepsilon _1$ and $\varepsilon _2$. If we do this in equation \ref
  {eq:num_29} and follow through the analysis that led
  to \ref {eq:L_resonance} we find that the
   resonance condition, using the Drude formula for the permittivity of the
  metal, \ref {eq:num_aa}, is $\omega =\omega _P\sqrt {2/3}$. \cite {Natta_SSC_1969_7_823}\cite
  {Coyle_PRL_2001_87_176801}

\bibitem{Natta_SSC_1969_7_823}
M.~Natta, ``Surface plasma oscillations in bubble"
\newblock Solid State Comm., {\bf 7}, 823-825 (1969).

\bibitem{Coyle_PRL_2001_87_176801}
S.~Coyle et~al., ``Confined Plasmons in Metallic Nanocavities"
\newblock Phys. Rev. Lett. {\bf 87}, 176801--4 (2001).

\bibitem{PandP}
W.~K.~H. Panofsky and M.~Philips,
\newblock {\em {Electromagnetism}},
\newblock (Addison-Wesley, 2nd edition, 1955).

\bibitem{AandM}
N.~W. Aschcroft and N.~D. Mermin,
\newblock {\em {Solid State Physics}},
\newblock (Holt Rinehart and Winston, 1st edition, 1976).

\bibitem{Fox_OPS}
M.~Fox,
\newblock {\em {Optical Properties of Solids}},
\newblock (Oxford University Press, Oxford, 2nd edition, 2010).

\bibitem{Note5}
Note that the lowest symmetry object for which the quasistatic approach is
  valid (i.e. that satisfies the requirement that the field in the particle be
  homogeneous) is an ellipsoid. \cite {Kang_ARMA_2008_188_93}

\bibitem{Kang_ARMA_2008_188_93}
H. Kang and G.~W. Milton, ``Solutions to the P\' olya--Szeg\"o conjecture and the weak Eshelby conjecture"
\newblock Archives of Rational Mechanical Analysis, {\bf 188}, 93-116 (2008).

\bibitem{NandH}
L.~Novotny and B.~Hecht,
\newblock {\em {Principles of Nano-Optics}},
\newblock (Cambridge University Press, 1st edition, 2006).

\bibitem{Myroshnychenko_ChemSocRev_2008_37_1792}
V.~Myroshnychenko et~al., ``Modelling the optical response of gold nanoparticles"
\newblock Chem. Soc. Rev. {\bf 37}, 1792-1805 (2008).

\bibitem{Note6}
Actually, the resonance condition can be met, but only if the surrounding
  medium has a complex permittivity; the surrounding medium will need to
  support gain, i.e. be capable of amplifying \cite {Lawandy_APL_2004_85_5040} so
  as to offset the damping, both radiative and non-radiative, of the metallic
  particle.

\bibitem{Lawandy_APL_2004_85_5040}
N.~Lawandy, ``Localized surface plasmon singularities in amplifying media"
\newblock App. Phys. Lett. {\bf 85}, 5040-5042 (2004).

\bibitem{Whitney_JPCB_2005_109_20522}
A.~V. Whitney et~al., ``Localized surface plasmon resonance nanosensor: a high-resolution distance-dependence study using atomic layer deposition"
\newblock J. Phys. Chem. B {\bf 109}, 20522-1--8 (2005).

\bibitem{Raschke_NL_2003_3_935}
G.~Raschke et~al., ``Biomolecular Recognition Based on Single Gold Nanoparticle Light Scattering"
\newblock Nano Lett. {\bf 3}, 935-938 (2003).

\bibitem{Kelly_JPCB_2003_107_668}
K.~L. Kelly, E.~Coronado, L.~L. Zhao, and G.~C. Schatz, ``The optical properties of metal nanoparticles: The influence of size, shape, and dielectric environment"
\newblock J. Phys. Chem. B {\bf 107}, 668-677 (2003).

\bibitem{Meier_OL_1983_8_581}
M.~Meier and A.~Wokaun, ``Enhanced fields on large metal particles: dynamic depolarization"
\newblock Opt. Lett. {\bf 8}, 581-583 (1983).

\bibitem{Kuwata_APL_2003_83_4625}
H.~Kuwata, H.~Tamaru, K.~Esumi, and K.~Miyano, ``Resonant light scattering from metal nanoparticles: Practical analysis beyond Rayleigh approximation"
\newblock Appl. Phys. Lett. {\bf 83}, 4625-4627 (2003).

\bibitem{Moroz_JOSAB_2009_26_517}
A.~Moroz, ``Depolarization field of spheroidal particles"
\newblock J. Opt. Soc. Am. B {\bf 26}, 517-527 (2009).

\bibitem{Tcherniak_NL_2010_10_1398}
A.~Tcherniak, J.~W. Ha, L.~S. Slaughter, and S.~Link, ``Probing a century old prediction one plasmonic particle at a time"
\newblock Nano Lett. {\bf 10}, 1398-1404 (2010).

\bibitem{Burrows_OE_2010_18_3187}
C.~P. Burrows and W.~L. Barnes, ``Large spectral extinction due to overlap of dipolar and quadrupolar plasmonic modes of metallic nanoparticles in arrays"
\newblock Opt. Express {\bf 18}, 3187-3198 (2010).

\bibitem{Knight_NL_2009_9_2188}
M.~W. Knight, Y.~Wu, J.~B. Lassiter, P.~Nordlander, and N.~J. Halas, ``Substrates matter: influence of an adjacent dielectric on an individual plasmonic nanoparticle"
\newblock Nano Lett. {\bf 9}, 2188-2192 (2009).

\bibitem{Rechberger_OC_2003_220_137}
W.~Rechberger et~al., ``Optical properties of two interacting gold nanoparticles"
\newblock Opt. Comm. {\bf 220}, 137-141 (2003).

\bibitem{Abajo_RMP_2007_79_1267}
F.~J. {Garc{\'{\i}}a de Abajo}, ``Colloquium: Light scattering by particle and hole arrays"
\newblock Rev. Mod. Phys. {\bf 79}, 1267-1290 (2007).

\bibitem{Hohenester_CompPhysComm_2012_183_370}
U.~Hohenester and A.~Tr{\"{u}}gler, ``MNPBEM - A Matlab toolbox for the simulation of plasmonic nanoparticles"
\newblock Comp. Phys. Comm. {\bf 183}, 370-381 (2012).

\bibitem{Parsons_JMO_2010_57_356}
J.~Parsons, C.~P. Burrows, J.~R. Sambles, and W.~L. Barnes, ``A comparison of techniques used to simulate the scattering of electromagnetic radiation by metallic nanostructures"
\newblock J. Mod. Opt. {\bf 57}, 356-365 (2010).

\bibitem{Noguez_JPCC_2007_111_3806}
C.~Noguez, ``Surface plasmons on metal nanoparticles: the influence of shape and physical environment"
\newblock J. Phys. Chem. C {\bf 111}, 3806-3819 (2007).

\bibitem{KandV}
U.~Kreibig and M.~Vollmer,
\newblock {\em {Optical properties of metallic clusters}},
\newblock (Springer, 1995).

\bibitem{Barnett_JPhysB_1996_29_3763}
S.~M. Barnett, B.~Huttner, and R.~Loudon, ``Decay of excited atoms in absorbing dielectrics"
\newblock J. Phys. B: Atomic Mol. and Opt. Physics {\bf 29},
  3763-3781 (1996).

\bibitem{LCandL}
P.~Lorrain, D.~R. Corson, and F.~Lorrain,
\newblock {\em {Electromagnetic Fields and Waves}},
\newblock (W. H. Freeman, 3rd edition, 1988).

\end{thebibliography}
\end{document}